\documentclass[preprint,12pt]{elsarticle}

 \usepackage{graphicx}

\usepackage{amssymb}
\newcommand\WINHAC[0] {\textsf{WINHAC}}

\newcommand\LHAPDF[0] {LHAPDF}

\newcommand{\Wp}     {{W^+}}

\newcommand{\Wm}     {{W^-}}   

\newcommand{\MZ}     {{M_Z}}

\newcommand{\qbar}    {{\bar q}}

\def\slashii#1{\setbox0=\hbox{$#1$}            
  \dimen0=\wd0                                 
  \setbox1=\hbox{\sl/} \dimen1=\wd1            
  \ifdim\dimen0>\dimen1                        
     \rlap{\hbox to \dimen0{\hfil\sl/\hfil}}   
     #1                                        
  \else                                        
     \rlap{\hbox to \dimen1{\hfil$#1$\hfil}}   
     \hbox{\sl/}                               
  \fi}

\journal{Acta Physics Polonica B}

\begin{document}

\begin{frontmatter}



%
\title{The LHC excess of four-lepton events interpreted as Higgs-boson signal: \\
background from Double Drell--Yan process?$^{\star}$}



\author[1]{Mieczyslaw Witold Krasny}
\author[2]{Wies{\l}aw P{\l}aczek}
\address[1]{Laboratoire de Physique Nucl\'eaire et des Hautes Energies, \\
          Universit\'e Pierre et Marie Curie -- Paris 6, Universit\'e Paris Diderot -- Paris 7, \\
          CNRS--IN2P3, 4 pl. Jussieu, 75005 Paris, France.}
\address[2]{Marian Smoluchowski Institute of Physics, Jagiellonian University,\\
         ul.\ Reymonta 4, 30-059 Krak\'ow, Poland.}
\begin{abstract}
 
We construct   a simple model of the Double Drell--Yan Process (DDYP) for proton--proton collisions 
and  investigate its possible contribution to the background for the Higgs-boson searches at the LHC.  
We demonstrate that  under the assumption  of the predominance of short range, 
${\cal O}(0.1)\,$fm, transverse-plane correlations of quark--antiquark pairs within the proton, this contribution
becomes important and may even explain the observed excess of the four-lepton events  at the LHC -- the 
events interpreted as originating from 
the Higgs-boson decays: $ H \rightarrow  ZZ^{*}  \rightarrow   4l$ and $ H \rightarrow  WW^{*}  \rightarrow   2l 2 \nu$.  

\end{abstract}


\end{frontmatter}

\vspace{35mm}
\footnoterule
\noindent
{\footnotesize
$^{\star}$The work is partly supported by the Programme of the French--Polish 
Cooperation between IN2P3 and COPIN no.\ 05-116,
and by the Polish National Centre of Science grant no.\ DEC-2011/03/B/ST2/00220.
}


\section{Introduction}
\label{Introduction}

The recent observation \cite{ATLAS_Higgs,CMS_Higgs} of an excess of events containing  a pair of photons,  or
opposite charge leptons associated with missing transverse energy,  or two pairs of opposite charge leptons,  in the region 
of invariant mass of ~125 GeV, has been readily interpreted as a discovery of a new Higgs-like boson. The 
characteristics of these events have been found to be well described by the present event generators 
which,  on top of the Standard Model (SM) background processes,  include the processes of production and decay of the Higgs 
boson  \cite{ATLAS_Higgs, CMS_Higgs}.

The road  to verify  if indeed the signal of  the Higgs boson is observed,  being pursued now by the CMS and  ATLAS 
experiments,  is to measure the coupling strengths, spin and the parity of the particle,  believed to be the source of the 
of excess events.  These tests are of indisputable importance. 
In our view they should, however,  be complemented by the experimental exclusion of other mechanisms which may contribute
to the observed excess of events,  including those which have not been, so far, implemented in the current
Monte Carlo generators of the background SM processes due to the lack of an adequate theoretical framework,  or the lack of 
input information, or both.

The mechanisms to be considered must not be in conflict with the present LHC $\Wp \Wm$ and $ZZ$ production data. 
The $\Wp \Wm$ and $ZZ$ cross sections measured by the CMS  \cite{CMS_cross_sections} and 
ATLAS \cite{ATLAS_cross_sections} collaborations are consistently above the theoretical predictions
allowing for a possible presence of such mechanisms. It has to be stressed, however, that   the statistical 
significance of the observed excess is weak  and the experiments claim the consistency of the measured and predicted 
values. 
In the following, we shall assume that any additional mechanism producing the four-lepton events must not increase the overall 
predicted cross-section for  the $\Wp \Wm$ and $ZZ$ on-shell pairs above the measured values.

In this paper, we propose a simple Charged Current (CC) and Neutral Current (NC) Double Drell--Yan Process (DDYP) model. 
We  use this model to investigate  a possible,  complementary mechanism producing  four-lepton events and 
demonstrate  that the DDYP may contribute to, and even explain  the Higgs-boson signals in the four-lepton channels
presented in \cite{ATLAS_Higgs} and  \cite{CMS_Higgs}.  Finally, we 
discuss the perspective of discriminating  between the above two mechanisms on the experimental ground.

The DDYP has  not, so far,  been  reported as a  potential 
source of the observed excess of events or even as a contribution to the SM background in the Higgs particle searches. This work is, 
to our best knowledge,  a first step in this direction. 

\section{Double Drell-Yan process}
\label{DDYP}


The phenomenological parton-model description of the DDYP process  was developed long time ago,
see e.g. \cite{Goebel}. 
We follow closely the phenomenological model of  \cite{Goebel} and 
explain below our modifications of  this model, and its technical 
implementation within a Monte Carlo event generator.
In the following, we shall ignore the DDYP involving two different protons of each of 
the colliding particle bunches. We have estimated  their contribution to be  below the level of 0.1 event 
for each $1\,$fb$^{-1}$ of the collected luminosity, i.e. negligible. 

Our departure point is the canonical,  factorized form of the cross section for double Drell--Yan process: 
\begin{eqnarray}
\sigma^{\rm DDYP}(p_1,p_2,p_3,p_4) = \frac{\sigma^{\rm SDYP}(p_1,p_2) \,\sigma^{\rm SDYP}(p_3,p_4 )}{S_{qq}}, 
\label{master_formula} 
\end{eqnarray}
where  
$p_i$ are the four-momenta of outgoing leptons (for each allowed $e$, $\mu$, $\nu_{e}$ and $\nu_{\mu}$ combinations), 
$\sigma^{\rm SDYP}$ is the Single Drell--Yan 
Process (SDYP) cross section and  $S_{qq}$ can be interpreted as  an effective transverse-plane correlation 
area over which the majority of double-quark annihilations
take place. This quantity is  denoted frequently in the literature as $\sigma_{\rm eff}$ or $S_{qq} = \pi R^2$, 
where $R$ is the equivalent radius of the circle having the surface equal to $S_{qq}$ \cite{Goebel}.

This factorized form assumes that the product of double-parton distribution functions:
$D_q(x_1,x_2, m_1^2,m_2^2) \times D_q(x_3,x_4, m_1^2,m_2^2)$, 
where $x_i$ are the fractional longitudinal momenta of annihilating partons, 
$m_1^2=(p_1+p_3)^2$,  and $m_2^2=(p_2+p_4)^2$, can be  written 
in form of a product of single-parton densities:
\begin{eqnarray}
D_q(x_1,x_2, m_1^2,m_2^2) D_q(x_3,x_4, m_1^2,m_2^2) = q(x_1,m_1^2)q(x_3,m_1^2)q(x_2,m_2^2)q(x_4,m_2^2). 
\end{eqnarray} 

We assume,  that  the $S_{qq}$  parameter is independent of the four-momenta 
of outgoing leptons within the region of the large lepton transverse momenta,  $p_T^l  \geq  5\,$GeV. 

There is  no experimental information constraining the value of the $S_{qq}$ parameter,
contrary to the analogous $S_{gg}$ and $S_{gq}$ parameters representing the transverse plane correlation area for 
processes in which at least one of the involved partons is a gluon. The  latter parameters 
were derived,  using  the factorized form of the Double-Parton Scattering (DPS) cross section, from the measured 
four-jet, three-jet and a photon, and two-jet and a $W$-boson cross-sections by the 
ISR \cite{ISR}, SppS \cite{SppS}, Tevatron \cite{Tevatron} and LHC \cite{LHC} experiments.  They are 
consistent with a simple model assuming a uniform density of uncorrelated gluons over the transverse area 
of the proton.

As a starting point to our studies, we have made a simple order-of-magni\-tude estimation 
and have found  that if  $S_{qq}  \approx S_{gg}, S_{gq}$,
the contribution of  the DDYP to the four-lepton production processes is sizeably  smaller than 
the Higgs-boson signal and can be,  to a good approximation, neglected. 
For this contribution to become non-negligible the following condition 
must be satisfied: $S_{qq}  \le 0.1 \times S_{gg}$. Can one  reject {\it a priori}  such a possibility?

There is no reason to expect that the transverse plane of both the quark--quark and antiquark--antiquark correlation 
areas are sizeably different  from  the gluon--quark and gluon--antiquark ones. This, however,  does not need to be the case for 
the quark--antiquark pairs.

In the region of $x<0.01$, relevant to this paper,  and at the $Q^2 \approx M_W^2, M_Z^2$ scale,   protons 
are composed almost entirely out of gluons. Their density outnumbers the density of sea quarks by a factor greater than $20$.  Conversions 
of gluons decelerated in the colour field of the bulk of the target partons, in the early stage (large Ioffe time \cite{Ioffe}),  
of the proton--proton collisions is the main source of the quark--antiquark pairs, in analogy to photon conversions in 
the effective electromagnetic field of atoms, rather than individual electrons,  being  the source of the electron--positron pairs.  
At the LHC beam energies the effective transverse plane size of the produced 
quark--antiquark colour and charge dipoles may be significantly smaller than the proton size and must be constrained  experimentally. 
This is especially true  for those of the vector bosons pairs that have all their  decay products detected in the fiducial 
volume of the LHC detectors. These pairs are produced by the quark--antiquark pairs which 
have, preferentially,  balanced longitudinal momenta. For such a configuration, the conversions of not only  the transversely 
but also longitudinally polarized virtual gluons must  be taken into account.

Motivated by the above considerations, we assume in the following  that  
the dominant contribution to the DDYP cross section comes from the process 
in which a $q\qbar$-pair coming from the sea of one of the 
colliding process annihilates with a $q\qbar$-pair coming from the sea of the second one. 

Assuming such a dominance, $\sigma^{\rm DDYP}(p_1,p_2,p_3,p_4)$ can be expressed as follows:
\begin{eqnarray}
\sigma^{\rm DDYP}(p_1,p_2,p_3,p_4) = 
\frac{\sigma^{q_s\qbar_s}(p_1,p_2) \sigma^{\qbar_sq_s}(p_3,p_4 )+ 
         \sigma^{\qbar_sq_s}(p_1,p_2) \sigma^{q_s\qbar_s}(p_3,p_4 )}{S_{q\qbar}}, 
\label{analysis_formula} 
\end{eqnarray}
where $\sigma^{q_s\qbar_s}$ ($\sigma^{\qbar_sq_s}$) is the cross section involving $q_s$ ($\qbar_s$) coming from the first proton and 
$\qbar_s$ ($q$) coming from the second one. 
The corresponding transverse plane correlation 
area $S_{q\qbar}$ of the $q\qbar$-pairs for the above process will be kept as a free parameter in our analysis, 
to be determined from comparisons of such a model to experimental data.

\section{Model implementation}
 \label{Model}

The numerical implementation of the DDYP model  is based on the Monte Carlo event generator 
\WINHAC~\cite{Placzek2003zg,Placzek:2009jy,WINHAC:MC}. It describes the single $W/Z$-boson production 
with leptonic decays  in hadronic collisions (proton--proton, proton--antiproton, proton--nucleus, nucleus--nucleus),
i.e. the CC and NC single Drell--Yan processes.   The parton-level hard processes are convoluted
with appropriate collinear PDFs taken from the \LHAPDF\ library \cite{Whalley:2005nh}. 
The current version of \WINHAC\ includes only the LO QCD matrix elements, however for
the $W$-boson processes it features the ${\cal O}(\alpha)$ YFS exclusive exponentiation for the electroweak corrections 
\cite{Placzek2003zg,Placzek:2009jy}. At this level, it has been cross-checked numerically to a high precision
with independent calculations  \cite{CarloniCalame:2004qw,Bardin:2008fn}.
In order to generate realistic event shapes, \WINHAC\  is interfaced with the {\sf PYTHIA~6.4}  generator \cite{Sjostrand:2006za} 
which performs the initial-state QCD (and QED) parton shower, appropriate proton-remnant treatment and
necessary hadronization/decays. This interface provides also improved generation of lepton transverse momenta
with respect to the original  {\sf PYTHIA6} program, which results in a good agreement with the NLO QCD predictions,
see \cite{Krasny:2012pa} for more details.

In the studies presented in this paper, we have used the {\sf MSTW2008NLO} parametrisation \cite{Martin:2009iq} of PDFs.
Since in our model only the processes involving the sea quarks are considered, the valence-quark PDFs have been set to zero. 
The total cross sections for the SDYP have been normalized to the ATLAS and CMS measured values. Thus, for the CC SDYP
in \WINHAC\ we have used the following values of the normalisation $K$-factor: $1.2618$ for ATLAS and $1.2840$ for CMS. 
In the case of  the NC SDYP, we have used the value of $1.26$ for both experiments.
 
In our simple model the double Drell--Yan processes are generated as two independent single Drell--Yan 
processes in which the quarks (antiquarks)  come from the opposite proton beams, 
i.e. the two quarks (antiquarks) in a DDYP event have the opposite longitudinal-momentum directions.
The longitudinal momenta of the quarks are  generated using the standard single-parton PDFs,  
while their transverse momenta are generated by the {\tt PYTHIA} generator which includes 
a Gaussian smearing of the primordial $k_T$ with $\sigma _{k_T} = 4\,$GeV. 
 
\section{Higgs-like signal of DDYP}
\label{Higgs}

\subsection{$ZZ$ channel}

\begin{figure}[h]
\begin{center}
\includegraphics[width=0.49\textwidth]{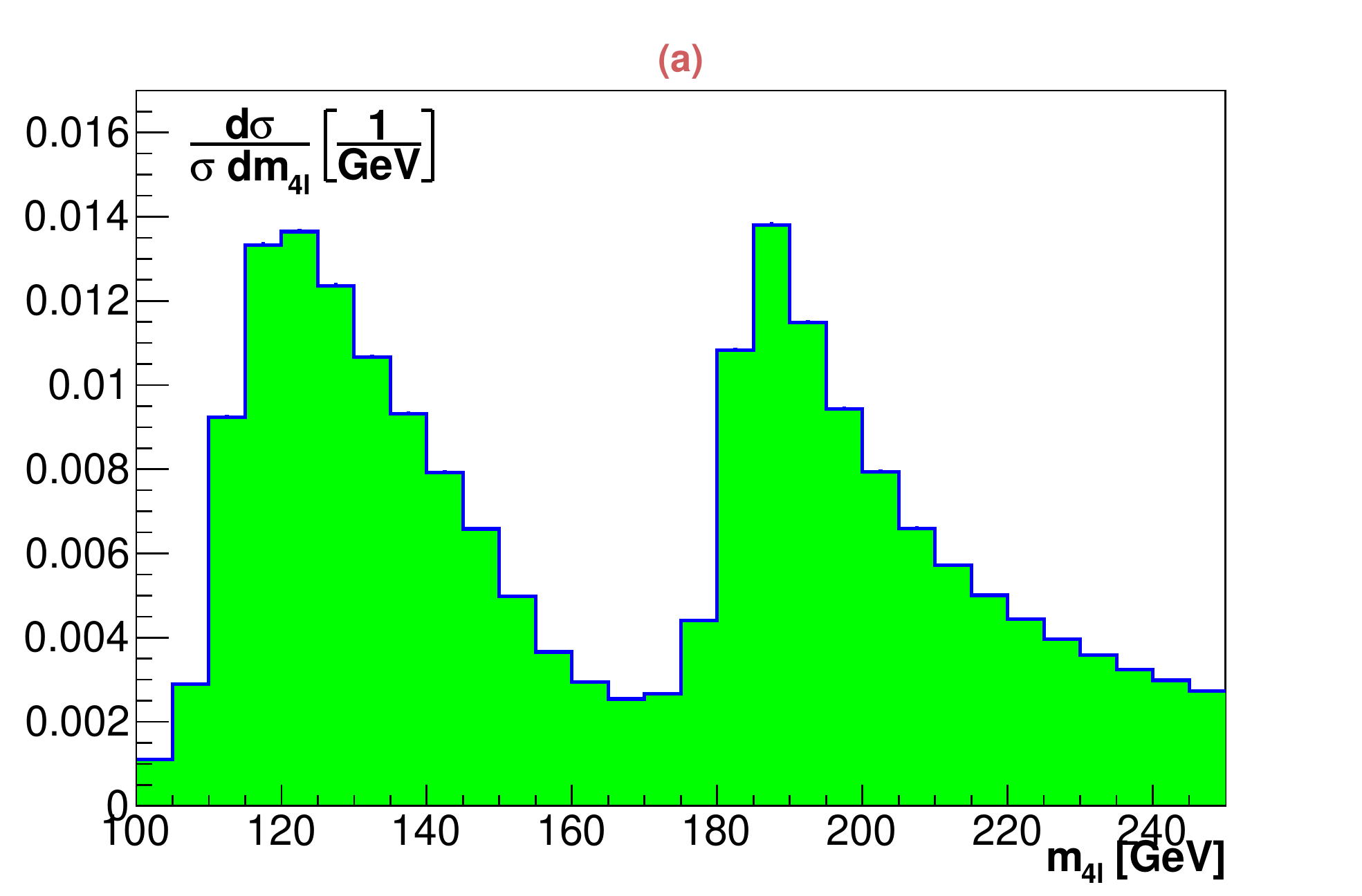} 
\includegraphics[width=0.49\textwidth]{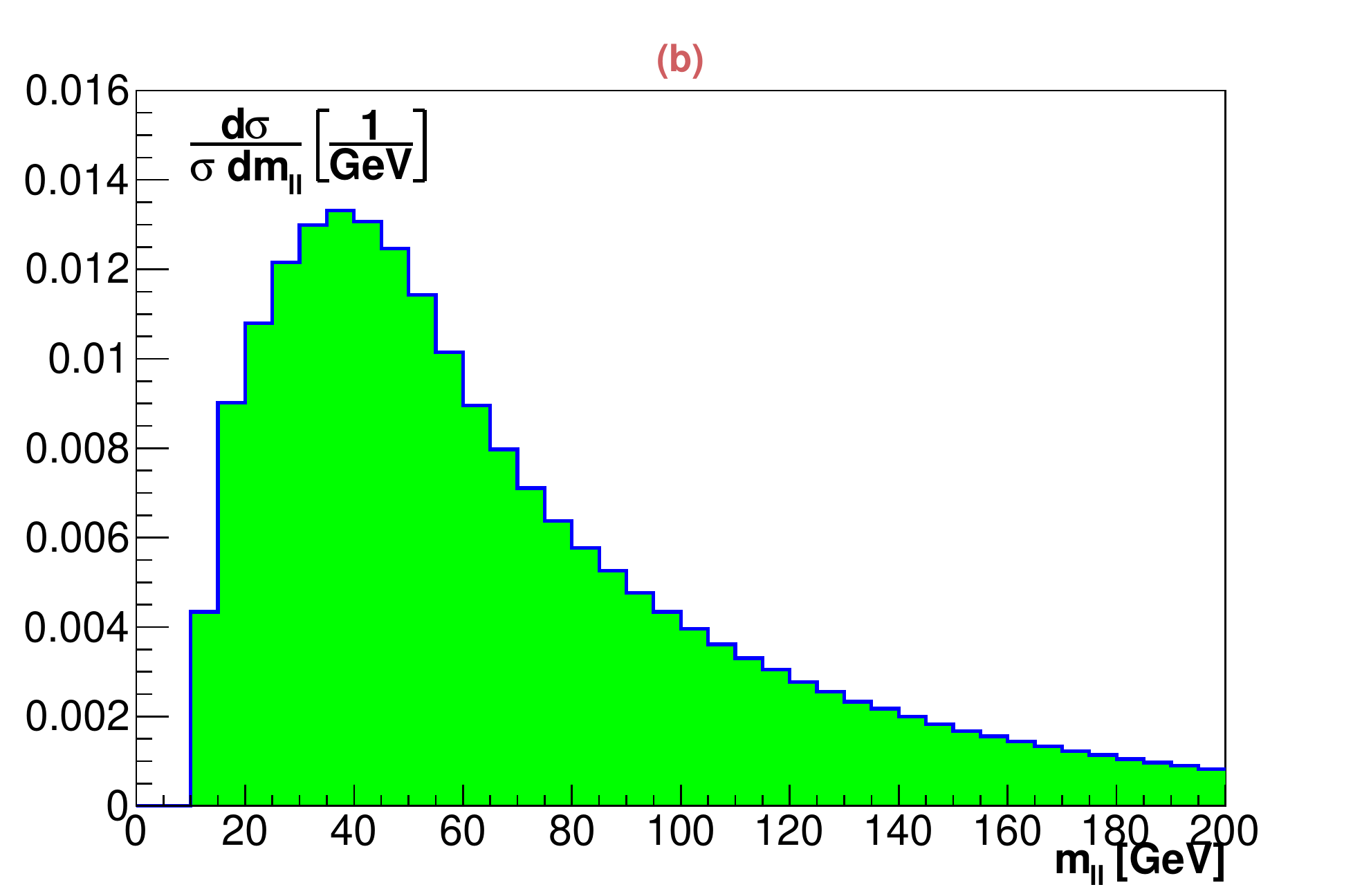} 
\end{center}
\caption{(a) The $4l$ invariant mass spectrum for the ATLAS event selection criteria \cite{ATLAS_Higgs};
(b) The $e\mu$ invariant mass spectrum for the CMS event selection criteria \cite{CMS_Higgs}.}
\label{4land2lmass}
\end{figure}

In Fig.~\ref{4land2lmass}a, we show the shape of the  four-charged-lepton invariant-mass distribution in our  
model of the NC DDYP process at the $8\,$TeV proton--proton collision energy. 
This plot includes the sum of the contributions coming from the following four charged lepton combinations: 
$\mu^+\mu^-\mu^+\mu^-$, $\mu^+\mu^-e^+e^-$, $e^+e^-\mu^+\mu^-$ and $e^+e^-e^+e^-$. The electron 
energies and the muon transverse momenta were smeared using the parametrized experimental resolution 
functions of the ATLAS detector \cite{ATLAS_detector}. The kinematical cuts on the electron and muon 
transverse momenta, pseudorapidities, invariant masses of the unlike-charge lepton pairs 
are the same as  the ones used in the  ATLAS paper \cite{ATLAS_Higgs}. The only cuts we could not 
implement are those corresponding to the detector-response related quantities, e.g. the 
lepton isolation cuts. 

Two peaks are clearly visible in this plot: the one  for $m_{4l} \approx 2 \MZ$  and the second 
one for $m_{4l} \approx 125\,$GeV where the excess 
of events was reported. While the appearance of the first peak, related to the DDYP production of 
a  pair of $Z$-bosons,  can be expected,  the appearance of the second one, in the mass region of the Higgs-like 
particle candidate,  is less trivial. Events in this region are characterized by  the invariant mass of the first 
of the opposite-charge lepton pairs, $m_{12}$,  in the $Z$-peak region, and the mass of the remaining pair, 
$m_{34}$, in the low mass  region. The  shape of the second peak 
is driven, for the masses smaller than the peak value, by the experimental cuts:  on the  minimal allowed 
transverse momenta of the leptons and on the minimal allowed $m_{34}$ values. For the masses between 
the peaks, the $m_{4l}$ distribution reflects the $ \approx 1/m_{l^+l^-}^{2}$ shape of the SDYP spectrum. 

Note that,  within the discussed model, there is no parameter which has been 
tuned to match the predicted and observed positions of the peak -- the peaking behaviour of the DDYP spectrum 
in the region where the excess of events was observed  is a generic feature of the DDYP and the event selection cuts,
largely independent of the approximations used in the construction of the model.  

The appearance of the second peak, around $125\,$GeV, of similar magnitude as the first one,  puts particular emphasis on 
the necessity of experimental verification of the importance of the DDYP mechanism. The excess of events 
in this region cannot, in our view, be (fully) attributed to the Higgs-like particle decays before 
investigating more closely the DDYP effects. Given the presence  of the peaking behaviour  in the DDYP spectrum, 
 the Higgs-boson hunting procedure can no longer be confined to a peak search, but must involve a detailed
 investigation of the excess-events properties. Since both the high-mass and low-mass peaks are of the same amplitude, 
 contrary to the canonical background peaking in the high-mass region only, any procedure 
 of extrapolation of the $4l$ spectra in the monitoring (high-mass) region to the Higgs-like particle signal (low-mass) region
 may become numerically unstable in the presence of the DDYP mechanism. 
 A small,  $\approx10$\%,  shift in the normalisation of the spectra 
 in the monitoring region is reflected by  a large, $\approx50$\%, shift  of the predicted background in the 
 Higgs signal region. 

\subsection{$WW$ channel}

In Fig.~\ref{4land2lmass}b, we show the shape of the  two-charged-lepton invariant mass, $m_{ll}$, distribution 
in our  model of the CC DDYP process for the proton--proton collision energy of $8\,$TeV. 
This plot shows the sum of the contributions for the $\mu^-e^+$ and $e^-\mu^+$ combinations, 
representing the highest signal-to-noise expectations for the Higgs-boson searches. The electron energies
and the muon transverse momenta have been smeared using the parametrized experimental 
resolution functions of the CMS detector \cite{CMS_detector}.  All the kinematical cuts of \cite{CMS_Higgs} 
have been implemented except, as before, the isolation cuts of leptons. The most notable  
difference with respect to the event selection and event reconstruction  procedures presented in 
\cite{CMS_Higgs} is the direct use of generated  four-momenta vectors of the neutrino and the anti-neutrino 
in the calculation of the missing transverse energy, $E_T^{miss}$,  the projected missing transverse energy, 
$E_T^{miss,proj}$ and the effective cut on the total transverse hadronic energy, $E_T^{had}$, which 
in our paper is used to approximate the selection conditions for the ``$0$-jet" events \cite{CMS_Higgs}.
 
 As in the case of the four-lepton invariant mass, this plot shows a significant fraction of events 
 in the low $m_{ll}$ region ($m_{ll} \leq  50\,$GeV) where the signal of the decay of the $125\,$GeV Higgs-like 
 particle is expected to show up \cite{CMS_Higgs}. Again, any attempt to attribute the excess events observed 
 in this region should, in our view, be preceded by the rejection of the DDYP mechanism as contributing 
 to the background estimation for the Higgs-boson searches.

\section{DDYP and Higgs-like particle evidence}
\label{fake_signal}

\subsection{$ZZ$ channel}

Apart from the excess of events in the $m_{4l}$ region of $120$--$130\,$GeV, the most striking feature 
of the $m_{4l}$ distribution presented in  \cite{ATLAS_Higgs} is that the predicted background tends 
to be lower than the data over the full mass region -- the ratio of the integrated numbers of the 
expected-to-observed events in the control region of $160\,$GeV $ < m_{4l} < 250\,$GeV estimated from 
the plot presented in \cite{ATLAS_Higgs} is 0.8 $\pm$ 0.08. Moreover, this ratio hardly  changes, to 
0.82 $\pm$ 0.07,  if the integration is made in the full range of the plotted masses%
\footnote{Unfortunately, the CMS Collaboration does not show 
the $ZZ$ peak in its paper \cite{CMS_Higgs},  therefore this trend cannot be verified directly using 
the CMS data. There is, however, an indication, coming from the CMS measurement of the total 
$ZZ$-pair cross section presented in \cite{CMS_cross_sections}, that the measured cross section 
is $\sim 10\%$ higher than the theoretical predictions.}:
$80\,$GeV $ < m_{4l} < 250\,$ GeV.

\begin{figure}[h]
\begin{center}
\includegraphics[width=1.0\textwidth]{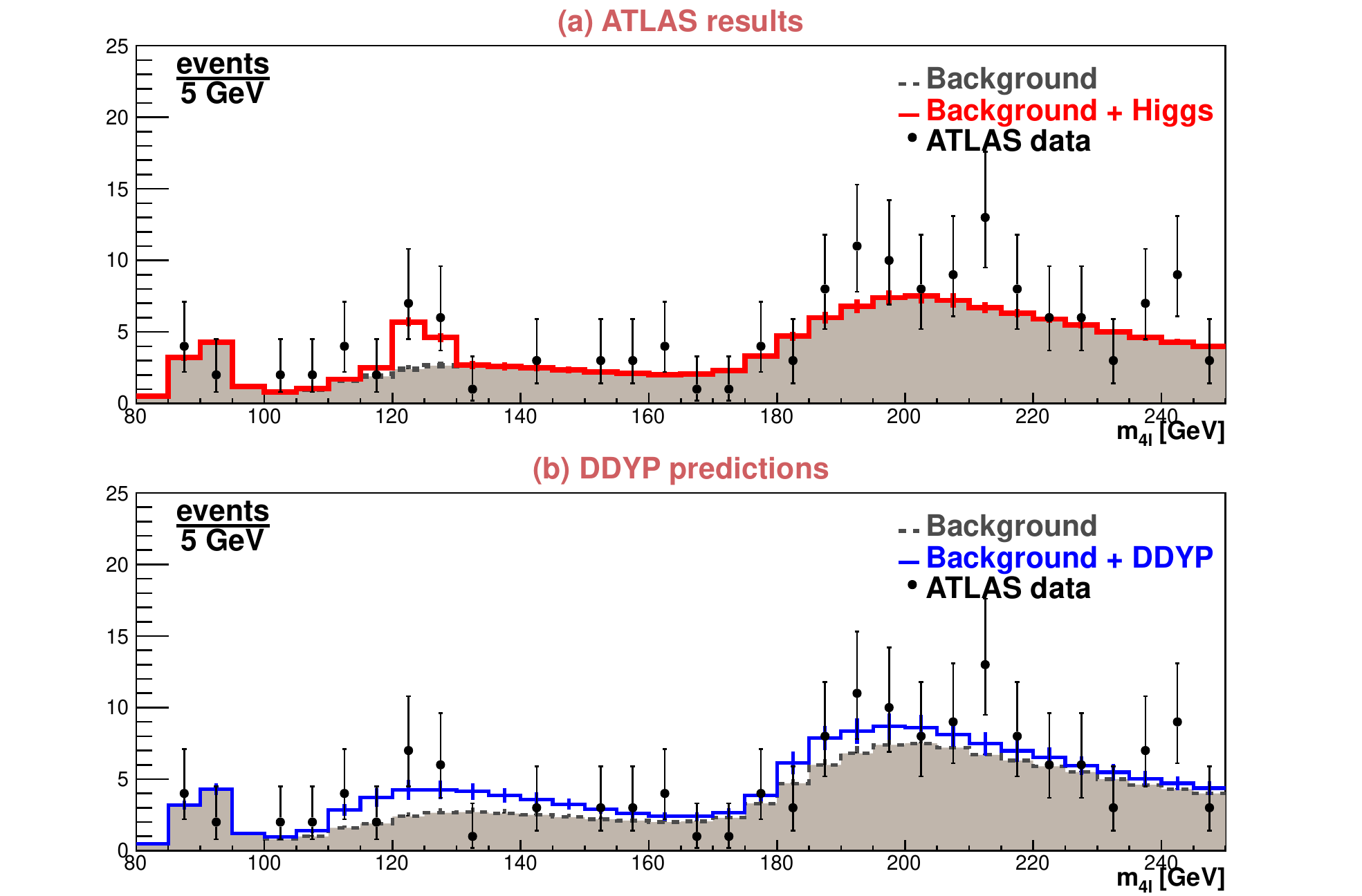} 
\end{center}
\caption{The $4l$ invariant-mass spectrum: (a) as presented by the ATLAS Collaboration in \cite{ATLAS_Higgs} 
and (b) with the Higgs-boson signal replaced by the DDYP contribution.}
\label{4lHiggs}
\end{figure}

The overall normalisation of the  DDYP contribution in our model cannot be predicted
and has  to be determined from the data. In the following,  we shall investigate what 
happens if the DDYP is added to the canonical background processes using the normalisation 
factor which equalizes the total number of predicted and observed events.

In Fig.~\ref{4lHiggs}, we compare the $m_{4l}$ plot presented in  \cite{ATLAS_Higgs}, in which 
the observed event distribution is compared to sum of the background  and the contribution of 
the Higgs-boson decays, with an  analogous plot in which, instead of the Higgs-boson 
contribution we have added the contribution of DDYP. The DDYP contribution 
was  normalized such that the integrals corresponding to
the data and to the background contributions presented in \cite{ATLAS_Higgs} plus the DDYP are equal. 
This plot has been obtained by merging two results: for the $7\,$TeV and 
$8\,$TeV proton--proton collision energies  with the respective  weights, 
representing the corresponding  fractions of the collected total luminosity. 
The quality of the overall fit of the data for  the ``background$+$Higgs" 
hypothesis, $\chi ^2 /{\rm dof} = 1.15\,$, is slightly worse than the one for the ``background$+$DDYP" hypothesis, 
$\chi^2/{\rm dof} = 1.04\,$, but both are equally acceptable; the corresponding $p_{0}$-values are respectively:
$p_{0} = 0.25\,$  and $p_{0} = 0.41\,$.

\subsection{$WW$ channel}

Both the ATLAS \cite{ATLAS_Higgs} and CMS experiments \cite{CMS_Higgs} observe 
an overall excess of  events in the $e\mu\nu\nu$--``$0$-jet" final state
with respect to the predicted background in the analysed kinematical region.  
In order to circumvent this mismatch, both experiments scaled up the background 
contribution in the kinematical region in which 
the decays of the $125\,$GeV Higgs boson do not contribute. They extrapolated subsequently 
the scaled-up spectra to the kinematical region where the Higgs decay contribution may show up, 
for more details, see  \cite{ATLAS_Higgs} and \cite{CMS_Higgs}. 

Since we do not have access to the unscaled distributions, we cannot construct, as in the case of the $ZZ$ channel, 
appropriate plots which would include the sum of the background determined in  \cite{ATLAS_Higgs,CMS_Higgs} 
and the DDYP contribution.
We can, however, use the measurement of the total $W^+W^-$ 
cross section at  $8\,$TeV, recently published by the CMS Collaboration \cite{CMS_cross_sections}, 
in order to evaluate  a possible contribution of the DDYP mechanism to the spectrum of the two-charged-lepton 
invariant mass, $m_{ll}$,   under 
the assumption that the inclusion of this process restores the agreement between the measured, 
$\sigma_{\rm exp}(W^+W^-) = 69.9 \pm 2.8\, {\rm (stat.)} \pm 5.6\, {\rm (syst.)} \pm 3.1\, {\rm (luminosity)}\,$pb, and predicted,   
$\sigma_{\rm th}( W^+W^-) = 57.3^{+2.4}_{-1.6}\,$pb, values of the cross section. 
It is intriguing to note that the ratio of the predicted to the observed cross sections, equal to 0.82 $\pm$ 0.08,
is the same as  the corresponding ratio for the $4l$ channel.  

The main difference of such a procedure, with respect to the one used in the previous section, 
is that one becomes sensitive to  all the detector-dependent contributions to the event-selection 
efficiency other than those related to kinematical cuts, for example  the lepton trigger and isolation 
efficiencies\footnote{In the case of the $4l$ channel, we rely only on the independence of the above efficiencies 
on the $m_{4l}$ value.}. Our predicted contribution of the DDYP, discussed below, represents thus its  upper 
limit. 

As shown in \cite{CMS_Higgs}, the decays of the $125\,$GeV Higgs particle contributes to the 
$m_{ll}$ spectrum mainly in the region $m_{ll}  < 50\,$ GeV. Using this result, we have estimated that in this region 
and for the integrated luminosity of $5.1\,$fb$^{-1}$ about $27\pm 5.4$  events are predicted to originate from 
the Higgs-boson decays. Our estimated  upper limit of the corresponding hypothetical contribution of the DDYP 
in this kinematical region, determined using our model normalized to the difference of the 
measured and the predicted $WW$ cross sections, is  $32 \pm 17$ events. 

Our conclusion from the above exercise is the same as before  for the $ZZ$ channel.
If we assume that the 2$\sigma$ excess of  the measured over predicted cross sections is not a statistical 
fluctuation but a real effect (this conjecture is  supported by the observation of a similar excess of events in both the 
$ZZ$ and $WW$ channels),  
the observed excess of events in the $m_{\rm ll}  < 50\,$GeV region can be attributed to the DDYP source.
Thus, again,  on a purely experimental ground, we are unable to 
 discriminate between the two hypotheses: (1)  that the excess of events in the $WW$ channel  is due to 
the Higgs-boson signal and (2) that the excess of events in the $WW$ channel is due to the DDYP contribution.


\section{Quark--antiquark  transverse-plane correlation area}
\label{S}

The quark--antiquark transverse plane correlation area,  $S_{q\qbar}$, is the only parameter of the 
model discussed in this paper. Its value  determines the overall normalisation  of the DDYP contribution. 

For the $WW$  final state,  $S_{q\qbar}$ is determined directly from absolute normalisation  
of the DDYP cross section to the difference between the measured and the predicted $W^+W^-$ cross 
sections by the CMS Collaboration \cite{CMS_cross_sections}. The resulting value is $S^{W^+W^-}_{q\qbar} = 0.075 \pm 0.04\,$mb.
 
For the $ZZ$ final state, we are bound to  make an assumption concerning the lepton-isolation 
efficiency, $\epsilon ^1 _{isol}$,  which is driven  by the detector-dependent cuts 
and, therefore,  cannot be implemented fully in our analysis. We have assumed a rather  conservative  allowed range 
for the ``per-charged-lepton" efficiency: $0.8 < \epsilon^1_{isol} <1.0$
to constrain the $S_{q\qbar}$ parameter by normalizing the calculated DDYP predictions 
to the difference of the predicted and the observed numbers of events in the $m_{4l}$ plot.  
The resulting value is $S^{ZZ}_{q\qbar} = 0.14 \pm 0.07\,$mb.

It is intriguing  to note that the above  values are compatible with the transverse size of the $W$ and $Z$ boson 
production zone, $\Delta r_T$,  determined from the Heisenberg uncertainty principle: $\Delta  k_T \Delta r_T \simeq 1$,
assuming $\Delta  k_T = 4\,$GeV -- the value used in the non-perturbative smearing of the  primordial transverse 
momentum of annihilating quarks in {\tt PYTHIA}. 

It is also very important to stress  that the average transverse plane correlation area of the quark--antiquark pairs $S_{q\qbar}$ 
must, within our model,  be a factor ${\cal O}(10)\,$  smaller  than  the corresponding ones for the gluon--gluon and gluon--quark  pairs,  
for the contribution of the DDYP to the four-lepton spectra to be non-negligible,  and a factor ${\cal O}(100)\,$  smaller 
for the full attribution of  the observed excess of events to the DDYP source\footnote{The above numbers may be modified  
in the case of presence of new initial or final-state interactions, specific to the colour-singlet, charge-neutral 
and spin-zero four-quark system, or to the $\Wp\Wm$ and  $ZZ$ boson pairs, produced at distances 
comparable to $1/ \Gamma_W, 1/ \Gamma_Z$.}.
The latter hypothesis, corresponding to the predominance of the short range, ${\cal O}(0.1)\,$fm 
transverse-plane correlations of the quark--antiquarks pairs within the proton, even if at the first sight unrealistic, should,   
in our view, be investigated  experimentally before it can be rejected.  

\section{Caveats} 
\label{Caveats}

The phenomenological framework which has been used in the presented above analysis is  obviously crude.
The DDYP  model is formulated using  the probabilistic language  rather than the  one based on quantum-mechanical amplitudes. 
It neglects the  colour, spin, flavour as well as longitudinal and transverse momenta correlations of the initial partons. They  
certainly play an  important role, in particular for the small $S_{q\qbar}$ region. It assumes that all 
the proton quarks and antiquark pairs are coming from the $S_{q\qbar}$ transverse-plane region.
Finally,  it neglects both the interference terms of the amplitudes of the DDYP processes with other 
sources of the vector-boson pairs, and in the case of $ZZ^*$ production the interference of 
diagrams with exchanged lines of the same flavour and charge leptons. 

A significant progress has been made recently in formulating the sound theoretical 
framework for a general description of  the Double-Parton Scattering (DPS) processes in the proton--proton collisions.
For a review of the recent progress, see \cite{Dokshitzer,Diehl} 
and the references quoted therein. One of the most important aspects of the present understanding 
of DPS, which is of critical relevance for the DDYP in  the small $S_{q\qbar}$ regime considered in this paper, and which was discussed 
in details  in \cite{Stirling},  is a correct theoretical handling of the DPS collinear singularity in the quark-loop integrals. 
The correct handling should  take into account the differences in the 
impact-parameter correlations of partons emerging from
perturbative  and non-perturbative processes and should avoid 
the double counting  of the single gluon--gluon scattering and the double-parton scattering contributions. While the former   
has already been taken into account in the evaluation of the background for the Higgs-boson signal in \cite{ATLAS_Higgs,CMS_Higgs}
by using the {\tt gg2ZZ}  \cite{gg2ZZ} and {\tt gg2WW} \cite{gg2WW} event generators, 
the latter has not  been, so far, reported  as a potential additional source of the four-lepton events at the LHC.
The importance of this point is amplified by the fact that,  in the selection of the Higgs-boson candidates decaying into pairs of 
$W$ and $Z$ bosons \cite{ATLAS_Higgs,CMS_Higgs},   neither a cut on the minimal absolute and relative transverse momenta 
of the corresponding lepton pairs,   nor a cut on the minimal
transverse momentum of the recoil hadronic system has been applied. The absence of such cuts exposes any future  calculation 
of the DDYP background for the Higgs-boson searches to the collinear and soft enhanced effects, discussed e.g.\ in \cite{Dokshitzer}.

The theoretical progress in the DPS framework has not been, so far, reflected  in development of  phenomenological 
tools allowing to study the DPS processes, in particular the DDYP, with realistic experimental  cuts. 
A corresponding DDYP event generator, using generalized two-parton distribution functions, taking into account 
the longitudinal-momentum as well as the transverse-impact-parameter correlations between the quark and the antiquark, and  including the 
colour, flavour and spin correlations simply does not exist. It is better, in our view, to construct  a crude model and investigate its consequences,  rather than to ignore {\em a priori} the DDYP mechanism altogether in the analysis of the background contribution to 
the Higgs-boson searches, for the reason that no precise technical tool exists. 
Obviously, the model presented in this paper cannot be more than  an initial  
tool for the investigation of the potential importance of the DDYP contribution to the Higgs-boson searches. 
A tool which  allows us  to define the  critical experimental tests to  discriminate between
the DDYP  and Higgs-boson mechanisms.  Being fully aware of all the theoretical caveats of the DDYP model 
presented in this paper, we shall  discuss in the next section how the hypothetical contribution of  the DDYP  
to the background of the Higgs-boson searches can  be rejected experimentally in a way which is the  least dependent 
on our  model approximations.

\section{Falsification of DDYP contribution}
\label{Falsification}

As  discussed in Section 5, on the basis of the data presented in \cite{ATLAS_Higgs} 
and~\cite{CMS_Higgs}
one cannot discriminate,  on purely experimental ground,   between the following  two  hypotheses: 

\begin{enumerate} 
\item
The observed excess of events in the $WW^*$ and $ZZ^*$ channels is entirely due to the production of the Higgs-like particle.
\item 
The excess events is  produced partially, or entirely by the DDYP mechanism.
\end{enumerate}

In order to define the critical experimental tests,  which may be used in the analysis of  the full statistics of events recorded till now at the LHC, we should first identify which event characteristics are similar in the two processes and which are distinct.  
 Let us first identify where the DDYP contribution will be hardly distinguishable from the Higgs-boson signal.

\begin{itemize}

\item 
The peaking behaviour in the four-lepton mass spectrum in the region of the observed excess of events:
in the case of the Higgs-boson signal the peak position reflects the Higgs-boson mass, whereas in the case of the DDYP it is driven both 
by the experimental cuts and by the remaining small  sensitivity to the assumption concerning the quark 
and antiquark distributions.

\item 
The DDYP, initiated by the annihilation of the two quark--antiquark dipoles, produces a pair of electroweak bosons of the preferentially 
opposite polarisations, mimicking perfectly  the decays of a scalar particle. 

\item 
If the gluons are the main origin of the small size quark--antiquark dipoles, 
the relative CM-energy dependence of  the DDYP effect and the Higgs-boson signal are 
similar, thus there is hardly any  gain from the method proposed in \cite{krasny} 
to distinguish between these two mechanisms on the basis of the measurement
 of the cross-section ratios at two different CM-energies of the proton--proton collisions.

\item 
The  DDYP process mimics  the custodial symmetry of the Higgs-boson coupling pattern 
to the $Z$ and $W$ bosons, in the sense that it generates  the same relative excess of events with respect to the 
corresponding canonical leading-twist SM background processes  for the $WW$ and $ZZ$ channels.
\end{itemize} 

There are,  however,  several characteristics of the DDYP events  which, independently 
of the approximations of the present DDYP model, allow reject the hypothesis that 
the DDYP  is the source of the observed excess of events. 

\begin{itemize}
\item 
The position of the peak for the Higgs-boson signal is invariant with respect to kinematical cuts, 
while the position of the DDYP peak  is cut-dependent. 
\item 
The excess of events in the region of  the $125\,$GeV peak  must be accompanied 
for the NC DDYP  by  the excess of events in the $m_{4l} > 2M_Z$ region
(for the $WW$ channel by the excess in the $m_{ll} > 80\,$GeV region).
Obviously,  the decays of the $125\,$GeV Higgs boson does not contribute to the high-mass regions.
\item 
The width of the $125\,$GeV peak,  in the case of the Higgs-boson signal,  is driven only by the detector 
experimental resolution which, for the $4l$ channel,  is of the order of $2\,$GeV for both experiments. 
In the DDYP case the peak is significantly broader -- its width  reflects both  by the experimental cuts 
and the $m_{l^+l^-}$ dependence of the SDYP cross section. 

\end{itemize} 

The above  differences are generic, i.e. largely independent of our  DDYP model approximations
and may be used in the experimental tests allowing to discriminate between the Higgs-boson and DDYP sources of the excess of events. 

As long as the LHC collaborations will not publish the distributions unfolded for the experimental effects, 
these tests can be done, in the fully quantitative way, only by the LHC collaborations. 

In the following we present, as an illustration,   a  concrete example of such tests: the study of the DDYP peak position as a function 
of the experimental cut on the minimal allowed transverse momentum of each of the leptons.  
The studies presented below cannot be directly compared to  the ATLAS and CMS data. 
They are made for an  ideal detector for which the selection efficiencies of isolated leptons are  
independent of their traverse momenta and of the lepton family.   

\begin{figure}[h]
\begin{center}
\includegraphics[width=1.0\textwidth]{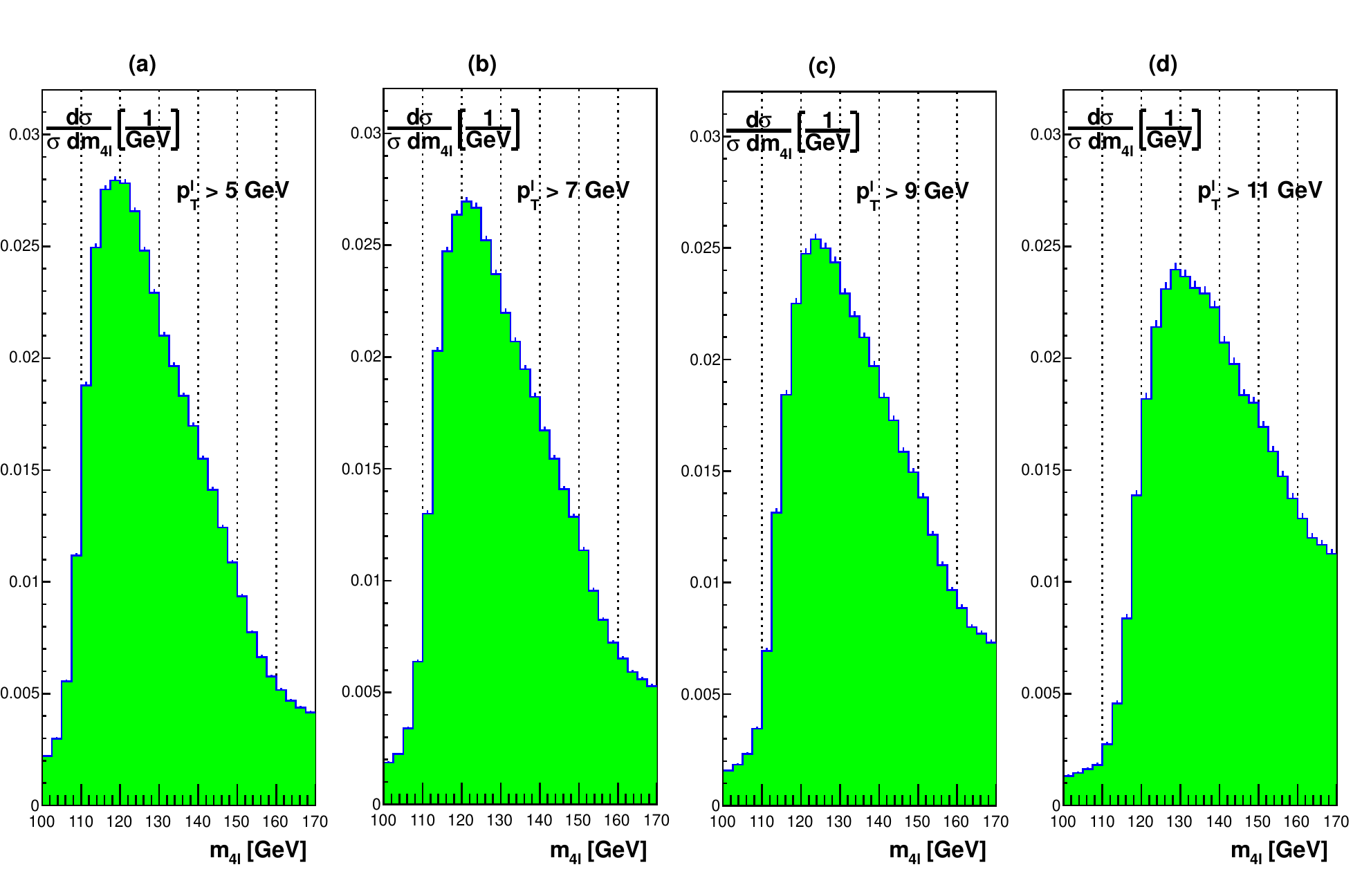} 
\end{center}
\caption{The $4l$ invariant-mass distributions for the following cuts on the minimal transverse momentum of each 
of the four charged leptons: (a) $5\,$GeV,  (b) $7\,$GeV, (c) $9\,$GeV and (d) $11\,$GeV.}
\label{Peak_position}
\end{figure}

In Fig.~\ref{Peak_position}, we show the evolution of the DDYP peak position as a function of the lepton minimal  $p^l_T$ cut for  the 
cut values of $5$, $7$, $9$ and $11\,$GeV. The peak positions were determined using the third-degree polynomial fit. They are, 
respectively: $119.1 \pm 1.3$, $121.1 \pm 1.3$, $124.2 \pm 1.3$, and $129.9 \pm 1.3\,$GeV. 
A significant variation of the peak position as a function of the  $p^l_T$ is observed. Its measurement,  using the unfolded  $m_{4l}$
distributions,  could put a stringent constraint on the importance of the DDYP contribution to 
the Higgs searches background. 

The present $p_T^l$ cut values which are implemented in the ATLAS and CMS Higgs searches \cite{ATLAS_Higgs,CMS_Higgs} are $7\, (6)\,$GeV for electrons (muons) in the ATLAS case, and $7\,(5)\,$GeV for electrons (muons) in the CMS  case. If the DDYP is the dominant source of the 
events in the $m_{4l}$ peak region then any observed differences in the peak positions for the ATLAS and CMS data, and  for the 4$\mu$  and 4e channels could be 
attributed to the respective differences in the $p^l_T$ cuts\footnote{Note that the $m_{\rm min}$
cut \cite{ATLAS_Higgs,CMS_Higgs}  is set in the above studies to $12\,$GeV -- the value implemented now by both the ATLAS and CMS
collaborations. We have checked that the peak positions, for the studied  range of the $p^l_T$ cuts, are stable with respect to adding the ATLAS specific linear rise of the $m_{\rm min}$ cut in the $m_{4l}$  range up to $190\,$GeV. We have also checked that the effective cut on the minimal allowed $p^l_T$ for the third lepton (if arranged in the order of decreasing $p^l_T$)  of $10\,$GeV,  
specific for the ATLAS selection 
criteria, affects only a negligible  fraction of the selected CMS events.}. 
In our DDYP model the peak position for the 4$\mu$ events is predicted to be shifted to a smaller  $m_{4l}$ value than that for the 
4$e$ events. Under the assumption of the  same $p^l_T$ dependence  of the electron and the  muon selection 
and isolation efficiencies, the predicted shift is $2\,$GeV  for the present CMS cuts and by a factor of $2$ smaller 
for the ATLAS ones.

\section{Conclusions and outlook}
\label{Conclusions}

In the presented paper, we have argued that on the basis of the published data the DDYP and Higgs-boson production 
mechanisms can not be discriminated -- they provide equally good description of the observed distributions. 
In order to assure,  beyond any doubt,  that the observed excess of events 
is entirely due to the Higgs-like boson decays, rather than the hypothetical DDYP contribution, 
the latter should be rejected on purely experimental grounds. 

If the  DDYP contribution is found to be non-negligible,  its detailed study 
could allow, for the first time,  to investigate experimentally the quark--antiquark correlations
within the proton and to make an important progress in  our basic understanding of  the relationship between the QCD and the proton structure. 
Strangely enough, 40 years after the QCD was proven to be the theory of the strong interactions,  the question: 
``What is the  underlying mechanism 
which correlates  the quark and antiquark longitudinal momenta and their transverse impact-plane positions within the proton
over the distances equal or smaller than the size of the proton?" 
cannot  be answered by the QCD and must be resolved experimentally. 

The DDYP provides not only the experimentally cleanest, but also the  most comprehensible environment
for such studies. 
The relative strength of the CC DDYP in the $\Wp\Wp +\Wm\Wm$ versus  $\Wp\Wm +\Wm\Wp$ 
processes could  resolve the transverse-plane correlations of the quark--quark pairs with respect 
to the quark--antiquark ones. Moreover, the relative strength of the DDYP effects in the proton--proton and 
proton--nucleus collisions could provide a crucial experimental insight into the relative 
importance of the short-distance (smaller that the nucleon size) and long-distance 
(comparable to the nucleon size) transverse-plane correlations of the quarks and antiquarks within 
nucleons. 



\end{document}